\documentclass[%
 aip,
 jmp,%
 amsmath,amssymb,
preprint,%
]{revtex4-1}

\usepackage{graphicx}
\usepackage{bm}
\usepackage{xcolor}
\usepackage[normalem]{ulem}

\begin{document}

\title{Gate-defined electron-hole double dots in bilayer graphene}

\author{L. Banszerus}
\email{luca.banszerus@rwth-aachen.de.}
\affiliation{JARA-FIT and 2nd Institute of Physics, RWTH Aachen University, 52074 Aachen, Germany,~EU}%
\affiliation{Peter Gr\"unberg Institute  (PGI-9), Forschungszentrum J\"ulich, 52425 J\"ulich,~Germany,~EU}
\author{B. Frohn}%
\affiliation{JARA-FIT and 2nd Institute of Physics, RWTH Aachen University, 52074 Aachen, Germany,~EU}%

\author{A. Epping}
\affiliation{JARA-FIT and 2nd Institute of Physics, RWTH Aachen University, 52074 Aachen, Germany,~EU}%
\affiliation{%
Peter Gr\"unberg Institute  (PGI-9), Forschungszentrum J\"ulich, 52425 J\"ulich,~Germany,~EU}%

\author{D. Neumaier}
\affiliation{
AMO GmbH, Gesellschaft f\"ur Angewandte Mikro- und Optoelektronik, 52074 Aachen, Germany,~EU}

\author{K. Watanabe}
\author{T. Taniguchi}
\affiliation{ 
National Institute for Materials Science, 1-1 Namiki, Tsukuba, 305-0044, Japan 
}%

\author{C. Stampfer}
\affiliation{JARA-FIT and 2nd Institute of Physics, RWTH Aachen University, 52074 Aachen, Germany,~EU}%
\affiliation{Peter Gr\"unberg Institute  (PGI-9), Forschungszentrum J\"ulich, 52425 J\"ulich,~Germany,~EU}%

\date{\today}

\begin{abstract}

We present gate--controlled single,  double, and triple dot operation in electrostatically gapped bilayer graphene. Thanks to the recent advancements in sample fabrication, which include the encapsulation of bilayer graphene in hexagonal boron nitride and the use of graphite gates, it has become possible to electrostatically confine carriers in bilayer graphene and to completely pinch-off current 
through quantum dot devices. Here, we discuss the operation and characterization of electron-hole double dots. We show a remarkable degree of control of our device, which allows the implementation of two different gate-defined electron--hole double--dot systems with very similar energy scales. In the single dot regime, we extract excited state energies and investigate their evolution in a parallel magnetic field, which is in agreement with a Zeeman-spin-splitting expected for a $g$-factor of two.

\end{abstract}


\maketitle

Graphene and bilayer graphene (BLG) are attractive platforms for spin qubits, thanks to their weak spin-orbit and hyperfine interaction, which promises long spin-coherence times~\cite{Trauzettel2007Feb}. This has motivated  substantial experimental efforts in studying quantum dot (QD) devices based on graphene~\cite{Ihn2010Mar,Ponomarenko2008Apr,liu10m,Guttinger2008Nov,Wang2010Dec,Volk2013Apr,Connolly2013May,Molitor2009Jun,Molitor2010Apr} and BLG~\cite{Volk2011Aug,Droscher2012Jul,all12,Goossens2012Aug}. The major challenge in this context is the missing band-gap in graphene, which does not allow confining electrons by means of electrostatic gates. A widely used approach to tackle this problem was 
to introduce a hard-wall confinement by physically etching the graphene sheet~\cite{Ihn2010Mar}. In this way, a number of important milestones, such as the implementation of charge detectors~\cite{Guttinger2008Nov,Wang2010Dec} and double quantum dots (DQDs)~\cite{Molitor2009Jun,Molitor2010Apr,Volk2011Aug} as well as the observation of the electron-hole crossover~\cite{gue09}, of spin-states~\cite{Guttinger2010Sep} and the measurement of charge relaxation times in graphene QDs~\cite{Volk2013Apr} have been reached. However, the influence of disorder, in particular the edge disorder~\cite{Bischoff2012Nov,Engels2013Aug}, turned out to be a major road block for obtaining clean QDs with a controlled number of electrons/holes and well tunable tunneling barriers. 

The problem of edge disorder can be completely circumvented in BLG, thanks to the fact that this material offers a tunable band-gap in the presence of a perpendicularly applied electric field~\cite{Zhang2009Jun,Oos2007Dec}, a feature that allows introducing electrostatic confinement in BLG. This route has been pursued by several groups to create QDs in BLG~\cite{all12,Goossens2012Aug,Eich2018Mar}. However, until very recently, essentially all devices were limited by leakage currents due to shortcomings in opening a clean and homogeneous band gap. A very recent breakthrough in this field has been the introduction of graphite back-gates~\cite{Li2017Oct,ove18j,Eich2018Mar}. Together with the technology of encapsulating BLG in hexagonal boron nitride (hBN), giving rise to high quality hBN-BLG-hBN heterostructures\cite{Engels2014Sep}, the use of a graphite back gate allows for a homogeneous and gate tunable band gap in BLG. This technological improvement allowed for an unprecedented quality of quantized conductance measurements\cite{ove18j} and, most  importantly, allowed realizing complete electrostatic current pinch-off. 
The latter finally offers the possibility of electrostatically confining carriers in BLG and to implement quantum dots with a high level of control and low disorder, as very recently demonstrated by Eich and coworkers\cite{Eich2018Mar}.

Here, we extend this approach and show the formation and operation of electrostatically defined single, double and triple dots in gapped BLG.

The device consists of a high-mobility hBN-BLG-hBN heterostructure fabricated by mechanical exfoliation and a dry van-der-Waals pick-up technique~\cite{Wang2013Nov,Engels2014Sep}. 
The BLG nature of the encapsulated flake is confirmed using confocal Raman microscopy\cite{Graf2007Feb} (data not shown). The heterostructure, consisting of a $t_1 \approx 30$~nm thick bottom and $t_2 \approx 20$~nm thick top hBN crystal, is placed on a roughly 10~nm thick exfoliated graphite flake serving as a local back-gate (BG). The use of the graphite BG improves the control over the device electrostatics by bringing the gate-structure substantially closer to the BLG layer than what is achievable with standard Si$^{++}$ back-gates (from typical values of ~300 nm to ~30~nm or less). At the same time, it helps screening potential fluctuations due to the Si/SiO$_2$ substrate. The BLG is electrically contacted by one-dimensional Cr/Au side contacts~\cite{Wang2013Nov}, following the strategy described in Ref.~\cite{Handschin2017Jan}. On top of the upper hBN crystal we place metallic split gates using electron beam lithography (EBL), evaporation of Cr/Au (5~nm/45~nm) and lift-off. The two split gates (SG) are 5~$\mu$m wide and are separated by approximately 250~nm (see scanning force microscope image in Fig.~1a). Next, we use atomic layer deposition (ALD) of alumina (Al$_2$O$_3$) to fabricate a 25~nm thick top-gate dielectric layer on the structure. In a last step, we fabricate three finger gates (FG1-FG3) crossing the trench defined by the split-gates using EBL and Cr/Au evaporation followed by lift-off. The finger gates have a width of around 200~nm and are separated by around 60~nm. A scanning force microscope image of the gate structure is shown in Fig.~1a and schematic cross-sections of the device are shown in Fig.~1b.

All measurements presented in this letter have been performed in a dilution refrigerator with 
a base temperature below 30~mK.
 We have measured the two-terminal conductance through the bilayer graphene by applying a symmetric DC bias voltage $V_b$ while measuring the current through the device.
In Fig.~1c we show the source-drain two-terminal resistance as function of the voltages applied to the BG and SG, with all FGs set to ground. The opening of a band gap in the BLG areas covered by the SG is observed as an 
enhanced resistance along the diagonal line in Fig.~1c with a (negative) slope of 0.8, in good agreement with the ratio of the BLG-BG and BLG-SG capacitances given by $t_2/t_1 \approx 0.7$.
From the data presented by Zhang \textit{et al.}\cite{Zhang2009Jun} we can estimate the size of the gate--induced band gap using the relation $E_g \approx 80$~meV$\cdot D$, where $D$ is the perpendicularly applied displacement field (in units [V/nm]). This approximation is valid for small displacement fields, $D < 0.4$~V/nm. Assuming that the charge neutrality point is exactly at $V_{\text{BG}}=V_{\text{SG}}=0$~V (which is in agreement with our data; see dashed lines in Fig.~1c), the displacement field can be estimated by\cite{Zhang2009Jun} $D \approx \epsilon_{\text{hBN}} [V_{\text{BG}}/(2t_1) - V_{\text{SG}}/(2 t_2)]$, where $\epsilon_{\text{hBN}} \approx 4$ is the dielectric constant of hBN.    
Fixing the BG and SG voltages at $V_{\text{BG}}=1.56$~V and $V_{\text{SG}}=-1.22$~V (see cross in Fig.~1c) corresponds to $E_g \approx$ 20~meV. In this configuration,  we start tuning the finger gates FG1 and FG2. Each of these gates allows to individually pinch-off transport through the channel defined by the SG (see e.g. Fig.~1d).  At sufficiently negative finger--gate voltages, the current shows signatures of quantized conductance (see steps at currents below 20 nA in Fig.~1d), which becomes unambiguously quantized steps in a four-terminal conductance measurement (see inset of Fig.~1d). These results are very similar to 
the work recently reported by Overweg et al.~\cite{ove18j}. At large negative finger--gate voltages, we reach maximum resistance values in the gigaohm regime.

In Fig.~2a we show the conductance as function of both $V_{\text{FG1}}$ and $V_{\text{FG2}}$. 
In the upper right corner (regime II) the finger gates are tuned such that there is no barrier along the channel (resulting in high conductance, $G >  e^2/h$). 
In regimes I and III one of the gates is tuned such that an island of holes is formed underneath 
gate FG1 (regime I) or FG2 (regime III), respectively. The formed p-island is tunnel-coupled to the n-doped channel leads giving rise to a single-dot configuration (see upper illustration in Fig.~2b). 
This behavior is well consistent with the very recent work by Eich et al.\cite{Eich2018Mar}. 
Fig.~2c shows a close-up of regime I (dashed box with arrow in Fig.~2a, but with less negative $V_{\text{FG2}}$)\footnote{Please note that the plot in Fig.~1d consists of only 200 data points making it hard to observe Coulomb peaks at very negative FG2 voltages.}. 
These data indicate (i) Coulomb peaks in the single-dot regime and (ii) the absence of any cross-talk due to FG2 on this dot (only vertical lines are observed in Fig.~2c). 
A line-cut at fixed FG2 of this plot is shown as inset in Fig.~2c.
From Coulomb diamond measurements we extract an addition energy of $E_a \approx 1.75$~meV (see white arrows in Fig.~1d and supplementary materials for details) and a gate lever arm of $\alpha=E_\text{a}/(e V_{\text{FG2}})$=0.05 (see Fig.~2d). Although this single-dot system is not perfectly clean (most likely  
because of some parallel dot close to the band edge caused by disorder in the 250~nm wide channel) we observe excited states  with a characteristic energy of $\Delta \approx 0.17$~meV (see right panel in Fig.~2d).

The electronic nature of the excited states can be shown by applying an in-plane magnetic field while measuring $dI/dV_b$ at finite bias. 
In Fig.~2e we show the evolution of a ground and excited state as function of an in-plane $B$-field, recorded at $V_b = 0.2$~mV (see e.g. dashed line in the right panel of Fig.~2d). 
These data show a zig-zag pattern due to a series of level crossings, which become apparent in the considered range of $B$-field  thanks to the small energy scale of the excited states. Notably, only two slopes can be detected in this data pattern, which only differ in sign and agree well with a Zeeman splitting $\Delta E_\text{Z} = g \mu_B B$ with $g=2$ (see black dashed lines). Here $\mu_B$ is the Bohr magneton. Taking $g=2$, we can extract  the energy of the the excited states from the relative position of the crossings. By repeating this type of measurements for different dot fillings, we extract a total of 14 excited state energies with a reasonably constant value of $\Delta =159 \pm$23~$\mu$eV. This value allows to estimate the diameter of the dot\cite{Volk2011Aug} by $d=(2\hbar^2/\Delta\,m^*)^{1/2}$, where $m^* = 0.033~m_{\text{e}}$ is the effective electron mass in BLG\cite{Kosh2006Jun} and $m_{\text{e}}$ is the electron mass. For the diameter of the island, we obtain $d \approx 170$~nm, which is compatible with our device geometry.

Tuning both finger gates to more negative voltage, we 
enter a regime that can be understood as a triple-dot configuration consisting of a p-island, tunnel coupled to an n-island, tunnel coupled to a second p-island, where the electric field-induced band gap serves as tunneling barriers (see lower illustration in Fig.~2b). In Fig.~2f  we show a close-up of Fig.~2a in this gate-voltage regime (see box (1) therein). These data show the typical pattern of the charge stability diagram of a triple-dot\cite{Gaudreau2006Jul}, with three characteristic slopes: (i) vertical features, related to the hole-dot discussed above, (ii) almost horizontal ones related to a hole dot (HD$'$) right underneath FG2, and (iii) features with  slope $-1$, which we attribute to an electron dot (ED) sitting right between FG1 and FG2.

Moving to even more negative values of $V_{\text{FG1}}$ and $V_{\text{FG2}}$, we finally reach the electron--hole double--dot regime (with a similar band alignment as in a nanotube electron-hole double dot\cite{Laird2013Jul}). In this regime, the negative voltage applied to the third finger gate FG3 ($V_{\text{FG3}}=-6$~V) 
breaks the  symmetry of the triple-dot lifting the right tunneling barrier of HD$'$. This is most likely due to an up shift of the band edge, which allows for a crossing of the charge neutrality point and the Fermi level in a region with a reduced band gap, resulting in an effectively ``non-gaped'' pn-junction (see illustration in Fig.~3a). The characteristic honeycomb pattern of the charge stability diagram\cite{Wiel2002Dec} of this double dot is shown in  Fig.~3b (measured at $V_b=0.1$~mV). A line-cut for fixed V$_{\text{FG1}}$ (dashed line in Fig.~3b) evidences the strong current suppression in the Coulomb blockaded regime, see Fig~3c. The current suppression become even more marked at lower bias voltages, see Fig.~3d (recorded at $V_b=-0.03$~mV). In this case,  elastic transport through the double dot is only possible when the electrochemical potential of both dots is aligned with the Fermi level in the leads, which is the case of the so-called triple-points at the corners of the honeycombs of constant charge.  At larger bias voltage, the triple points become triangular-shaped regions of finite conductance, see Fig.~3e (recorded at $V_b=0.17$~mV). The size of these regions allows to determine the conversion factors between gate--voltage and energy (see discussion below), while the broadening of the cotunneling lines provides insights on the bias voltage drop. From the substantially stronger broadening of the 
lines associated with the electron dot, we conclude that a large fraction of the bias voltage drops over ED.

This double dot regime shows a surprisingly high stability in finger--gates voltage space. Fig.~4a represents a large--range charge stability diagram of the double dot regime as function of $V_{\text{FG1}}$ and $V_{\text{FG2}}$ at a fixed bias of $V_b=70~\mu$V. 
The features associated with the co-tunneling lines show a remarkable periodicity over the entire presented voltage range of Fig.~4a that spans the addition of more than 100 holes and more than 100 electrons in each dot, indicating that the charging energy of the dots remains constant over this whole range. We verify these numbers by explicit counting of the charging events of the ED and the HD (see supplementary figure~4).
This periodicity is further confirmed by taking the two-dimensional Fourier transform of the data, see in Fig.~4b. Here, the diagonally spaced maxima are associated with the 
cotunneling lines related to the ED, which is located between FG1 and FG2, while the vertical features are associated with  the cotunneling lines related to the HD, which resides below FG1 and does not tune with FG2.
The $x$-- and $y$--component of the maximum highlighted by the circle (I) describe the cotunneling line spacing frequency with respect to FG2, $f^{\text{ED}}_{\text{FG2}} \approx 34$~V$^{-1}$, and FG1, $f^{\text{ED}}_{\text{FG1}} \approx 38$~V$^{-1}$, which nicely coincide with the cotunneling line spacing extracted from Fig.~3b, of around 33~mV and 25~mV  (see inset of Fig.~4a), respectively.
Similarly, the frequency of the vertical line marked by II, $f^{\text{HD}}_{\text{FG1}}  \approx 76~$V$^{-1}$, nicely matches  the spacing extracted from Fig.~3b of around 12~mV. These results are a good indication that this gate--defined double--dot system presents a parabolic confinement potential.

These results exhibit a remarkable degree of reproducibility.  In Fig. 5 we present data from a second double--dot system obtained by interchanging the role of FG1 and FG3, i.e. in this double dot the HD$'$ is located below FG3, while the ED$'$ sits between FG2 and FG3 (see illustration in Fig.~5b). Fig.~5b and Fig.~5c show high bias $V_b=0.17$~mV and low bias $V_{b}=-0.03$~mV charge stability diagrams of this double dot, which compare very well with those of the other configuration presented in Fig.~3, as one would expect from the geometry of the system. For a more quantitative comparison between the two double-dot systems, we use the charge stability diagrams at high bias (e.g.  V$_b=0.17$~mV as shown in Fig. 5e) to extract the relevant energy scales (see schematic in Fig.~5d and figure caption). From this kind of analysis we extract the total capacitance of the dots, which are $C_{\text{HD}}\,=\,297$~aF and $C_{\text{ED}}\,=\,369$~aF for the system of Fig. 3, and $C_{\text{HD}'} = 400$~aF and $C_{\text{ED}'} = 582$~aF   for the system of Fig. 5.
This corresponds to single-dot charging energies $E_{C}^{HD}=\,0.53$~meV and $E_{C}^{ED}\,=\,0.43$~meV for the first system, and  $E_{\text{C}}^{\text{HD}'} = 0.39$~meV and $E_{\text{C}}^{\text{ED}'} = 0.3$~meV for the second one. In both systems the electron dot has a roughly $0.1$~meV smaller charging energy compared to the hole dot located below a finger gate. The overall level of reproducibility of our double dot systems should be compared with the very high variability of etched dot systems, where edge roughness and fabrication contaminations played a large role in determining the behavior of each device.

In conclusion, we have showed single--, double-- and triple--dot operation of a gate-controlled BLG quantum dot device based on the electrostatic confinement of carriers. 
Our device is based on three finger gates, which allows to locally tune the band-gap and the chemical potential in a one-dimensional channel formed by two split gates. In this device geometry, the presence of the third finger gate seems to be essential to reach the double--dot regime, which is then formed by an electron dot tunnel coupled to a hole dot. This electron--hole double dot regime shows clear features of parabolic confinement over a remarkably wide range of gate voltages. Exchanging the role of the finger gates, we can study two different electron-hole double dot systems, which show extremely similar behavior, in agreement with the geometry of the system and our understanding of its behavior. This work represents a first step towards the experimental investigation of spin relaxation times in BLG quantum dots. Next necessary steps are the further down-scaling of the device--size, the implementation of charge detectors and of high frequency read out schemes, which are all realistic goals with the current technology.  Furthermore, thanks to the recent advancements in the synthesis of high quality BLG grown by chemical vapor deposition~\cite{Schmitz2017Jun}, which will allow for the synthesis of isotopically purified $^{12}$C BLG, this technology promises realizing spin-qubit systems with very long life times.

We thank M. M\"uller for help with the setup and F.~Hassler, F. Haupt and K. Ensslin for helpful discussions. Support by the ERC (GA-Nr. 280140),
the Helmholtz Nano Facility\cite{Albrecht2017May}, the DFG are gratefully acknowledged. This project has received funding from the European Union’s Horizon 2020 research and innovation programme under grant agreement No 785219. Growth of hexagonal boron
nitride crystals was supported by the Elemental Strategy Initiative conducted by the MEXT,
Japan and JSPS KAKENHI Grant Numbers JP26248061, JP15K21722 and JP25106006.

\bibliographystyle{apsrev4-1}

\bibliography{literature}

\newpage

\begin{figure}[t]
\centering
\includegraphics[draft=false,keepaspectratio=true,clip,width=0.45\linewidth]{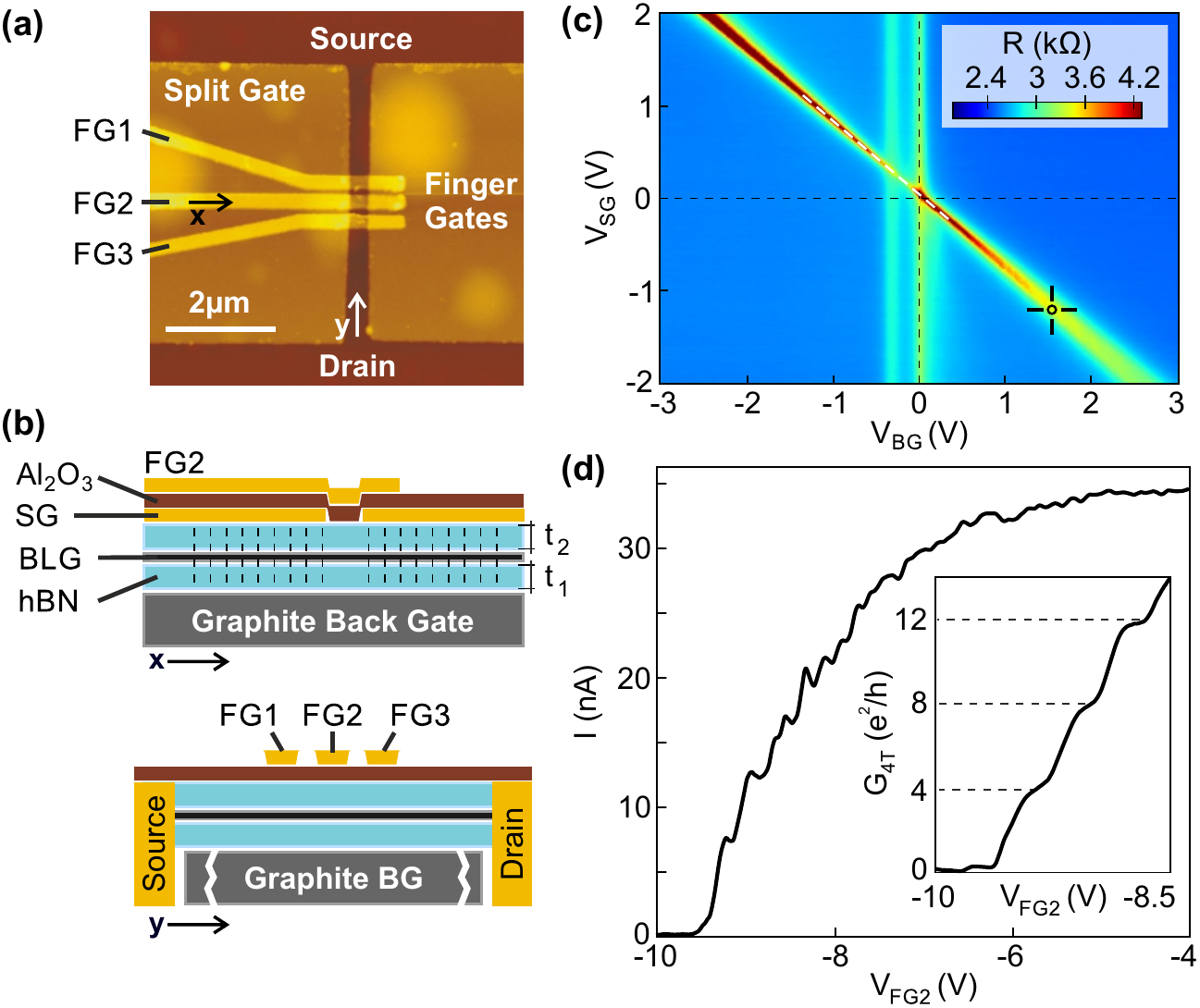}
 \caption[fig01]{(a) Scanning force microscopy image and schematic cross sections (b) of the device. 
For more information on labeling see text.
(c) Two terminal resistance of the device as function of  V$_{\text{BG}}$ and V$_{\text{SG}}$ with all FG voltages set to ground. (d) Current through the device as function of V$_{\text{FG2}}$ at a bias of $V_b=$~100~$\mu$V and fixed $V_{\text{BG}}$ and $V_{\text{SG}}$ indicated by the cross in panel c). Inset: Four-terminal measurement highlight the presence of quantized conductance (see dashed lines at 4, 8 and 12 $e^2/h$). Here, a serial resistance of 600~$\Omega$ has been subtracted to account for the 5~$\mu$m long channel. }
\label{fig01}
\end{figure}

\begin{figure*}[t]\centering
\includegraphics[draft=false,keepaspectratio=true,clip,width=\linewidth]{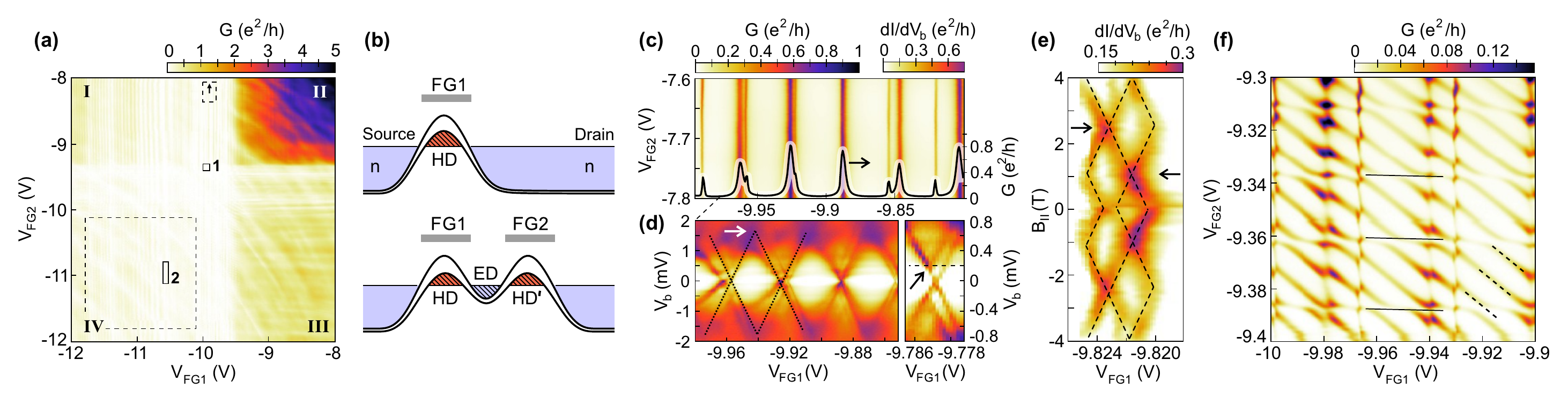}
\caption[fig02]{(a) Conductance as function of the finger gate voltages $V_{\text{FG1}}$ and $V_{\text{FG2}}$ at fixed BG and SG voltages (see cross in Fig.~1c) and $V_{\text{FG3}}=-6$~V. For more information on the different transport regimes marked by I--IV see text. 
(b) Schematic representations of the band alignment for the single and triple dot regime. (c) Charge stability diagram and Coulomb peaks for the single dot regime (I). The corresponding color bar is on the top left.
(d) Coulomb diamonds of the single dot regime. The right panel highlights the presence of excited states.
The corresponding color bar (for the main panel) is on the top right of panel c.
(e) Magneto-spectroscopy of excited states in the single dot regime. The black lines correspond to a g-factor of $g=2$, the arrows indicate the position of the level crossings. (f) Charge stability diagram of the triple dot regime. The black lines highlight the slopes corresponding to HD$'$ (solid line) and ED (dashed line). }
\label{fig02} 
\end{figure*}

\begin{figure}[tb]\centering
\includegraphics[draft=false,keepaspectratio=true,clip, width=0.45\linewidth] {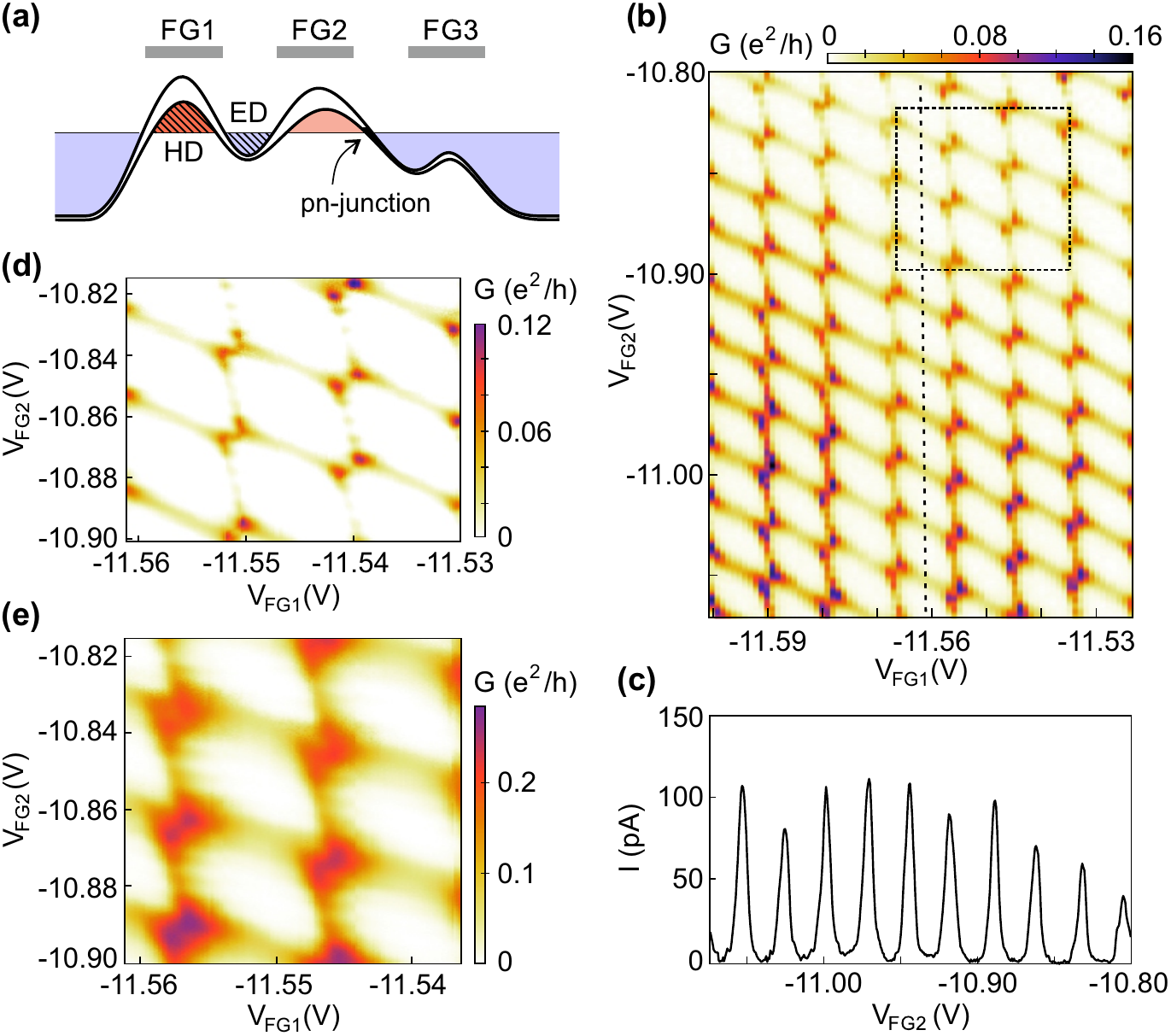}
\caption[fig03]{(a) Schematic representation of the band alignment in the double dot regime. While one HD is formed underneath FG1 the ED resides between two p-regions formed by FG1 and FG3. By detuning the right tunneling barrier of FG2 HD$'$ is destroyed (compare to Fig.~2b). (b) Charge stability diagram of the double dot as function of $V_{\text{FG1}}$ and $V_{\text{FG2}}$. 
(c) Current as function of V$_{\text{FG2}}$ at fixed V$_{\text{FG1}}$ (marked by the line in panel b).
(d) Close-up of the charge stability diagram (marked by the rectangle in panel b) recorded at $V_b=-0.03$~mV. (e) Charge stability diagram recorded at $V_b=0.17$~mV approximately in the same region as panel d. }
\label{fig03}
\end{figure}

\begin{figure}[t]\centering
\includegraphics[draft=false,keepaspectratio=true,clip,%
                   width=0.45\linewidth]%
                   {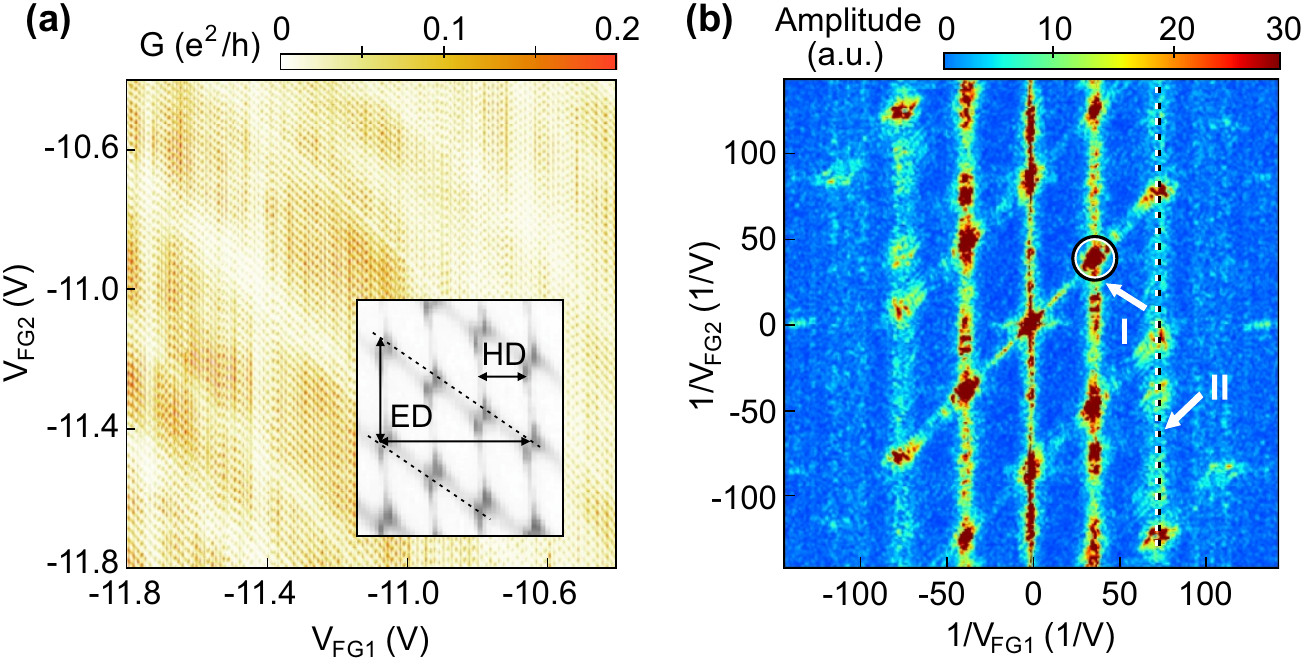}
\caption[fig05]{(a) Large--range charge stability diagram of the double--dot regime as function of  $V_{\text{FG1}}$ and $V_{\text{FG2}}$ (these data represent a close-up of the are enclosed by the large dashed rectangle Fig.~3b). The inset shows data from Fig.~3b and the spacing between the cotunneling lines. (b) Two-dimensional Fourier transform of the data presented in panel a. The diagonally oriented periodic features stems from the cotunneling lines associated with the ED, while the periodic vertical features  stems from cotunneling lines associated with the HD. The circle and the dashed line mark the features that exactly correspond to the line spacing extracted from the inset in panel a.     }
\label{fig04}
\end{figure}

\begin{figure}[tb]\centering
\includegraphics[draft=false,keepaspectratio=true,clip, width=0.45\linewidth] {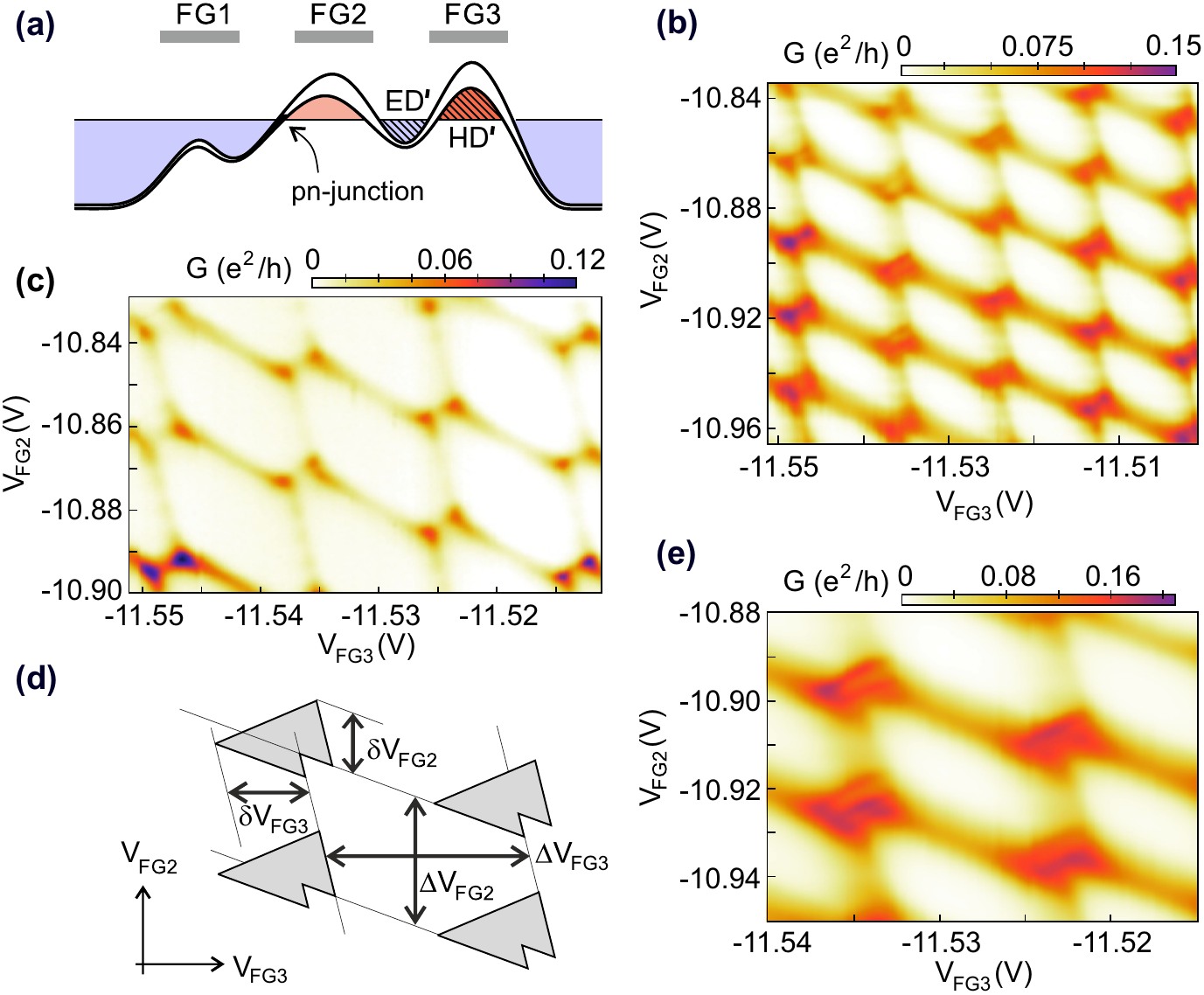}
\caption[fig04]{(a) Double dot configuration in which FG2 and FG3 are used to form an electron-hole double dot. (b, c) Bias--spectroscopy map recorded at $V_b=0.17$~mV and $V_{b}=-0.03$~mV, respectively. (d) Schematic representation of the charge stability measurements at high bias, indicating the relevant quantities for extracting the total capacitance and the charging energy of the dots.  For example, the total capacitance of the HD$'$ dot can be extracted as $C_{\text{HD'}}\,=\,C_{\text{FG}3}/\alpha_{\text{FG3}}$, where $C_{\text{FG}3}=e/\Delta V_{\text{FG3}}$ and $\alpha_{\text{FG3}}=V_b/\delta\,V_{\text{FG}3}$ are the capacitance and the lever arm between FG3 and the HD$'$, respectively. The single-dot charging energy is $E_{C}^{\rm HD'}=\alpha_{\text{FG3}}\,\Delta V_{\text{FG}3}$. Analogous relations hold for the ED$'$, with ED$'$ $\leftrightarrow$ HD$'$ and FG2 $\leftrightarrow$ FG3. Quantitatively we obtain  $\alpha_{\text{FG3}}=0.045$, $\alpha_{\text{FG2}}\,=\,0.013$, $C_{\text{FG3}}=13.4$~aF and $C_{\text{FG2}}=4.8$~aF. 
 (e) High-resolution zoom-in on four triple points measured at $V_{\text{b}}=0.17$~mV used to extract the values above.
In all the data present in this figure V$_{\text{FG1}}$ is set to $V_{\text{FG1}}=-7.15$~V. }
\label{fig05}
\end{figure}

\end{document}